\documentclass[
,prl
,groupedaddress
,showpacs
,amsmath
,amssymb
,a4paper
,twocolumn
]{revtex4}

\usepackage[hmarginratio=1:1,top=3cm,height=24.7cm,width=18cm,a4paper]{geometry}

\usepackage{graphicx}
\usepackage{dcolumn}
\usepackage{bm}

\begin{document}

\preprint{APS/123-QED}

\title{Electromagnetically induced transparency on a single artificial atom}

\author{A.~A.~Abdumalikov,~Jr.$^1$}
\email{abdumalikov@zc.jp.nec.com}
\altaffiliation[On leave from]{Physical-Technical Institute,
Tashkent 100012, Uzbekistan}
\author{O.~Astafiev$^{1,2}$}
\author{A.~M.~Zagoskin$^{3}$}
\author{Yu.~A.~Pashkin$^{1,2}$}
\altaffiliation[On leave from]{Lebedev Physical Institute, Moscow 119991, Russia}
\author{Y.~Nakamura$^{1,2}$}
\author{J.~S.~Tsai$^{1,2}$}
\affiliation{
$^{1}$RIKEN Advanced Science Institute, Wako, Saitama 351-0198, Japan\\
$^{2}$NEC Nano Electronics Research Laboratories, Tsukuba, Ibaraki 305-8501, Japan\\
$^{3}$Department of Physics, Loughborough University, Loughborough, LE11 3TU Leicestershire, UK
}

\date{\today}

\begin{abstract}
We present experimental observation of electromagnetically induced transparency (EIT) on a single macroscopic artificial ``atom'' (superconducting quantum system) coupled to open 1D space of a transmission line. Unlike in a optical media with many atoms, the single atom EIT in 1D space is revealed in suppression of reflection of electromagnetic waves, rather than absorption. The observed almost 100\% modulation of the reflection and transmission of propagating microwaves demonstrates full controllability of  individual artificial atoms and a possibility to manipulate the atomic states. The system can be used as a switchable mirror of microwaves and opens a good perspective for its applications in photonic quantum information processing and other fields.
\end{abstract}

\maketitle

\begin{figure}
\includegraphics{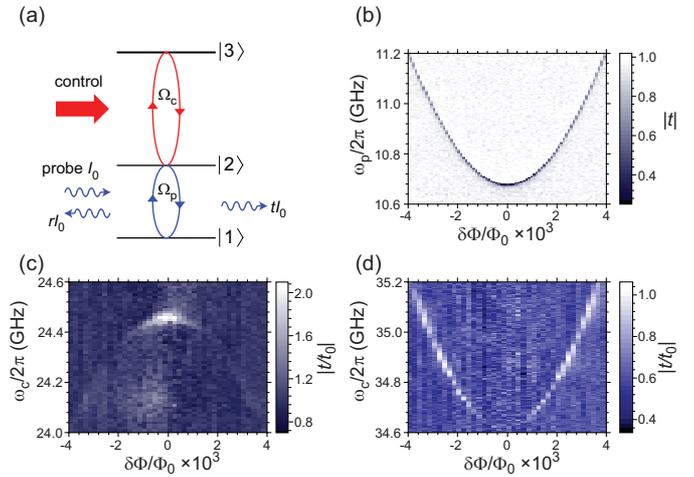}
\caption{\label{fig1} (Color online) Three-level cascade system. (a) Schematic of the three-level cascade atom used to produce electromagnetically induced transparency -- the physical phenomena for the quantum switch. Transmission of a weak propagating wave at the frequency $\omega_p$ close to the transition frequency $\omega_{21}$ is controlled by another field coupling states $|2\rangle$ and $|3\rangle$. The corresponding Rabi drive amplitudes are denoted by $\Omega_{\rm c}$ and $\Omega_{\rm p}$, respectively. (b) Transmission spectrum for $|1\rangle\leftrightarrow|2\rangle$ transition. (c) Spectrum of $|2\rangle\leftrightarrow|3\rangle$ transition. The spectroscopy signal appears only around the degeneracy point. The transition frequency between states $|2\rangle$ and $|3\rangle$ at the degeneracy ($\delta\Phi = 0$) is $\omega_{32}=24.465\,$GHz. (d) Spectrum of $|1\rangle\leftrightarrow|3\rangle$ transition. At the degeneracy point $\delta\Phi=0$ the transition is prohibited, and the spectroscopy signal vanishes indicating that the cascade system is realized.}
\end{figure}

Coherent evolution of atomic population under resonant coherent drive, known as Rabi oscillations in two-level atoms \cite{Scully}, leads to quantum interference phenomena in a more complicated case of a multi-level atom. In particular, in a three-level atom driven by two resonant waves (Fig.~\ref{fig1}a), the destructive interference between different excitation pathways cancels out population of one of the atomic states effectively turning the atom to the ``dark state''. This leads to the suppression of transition from the ``dark state'', revealed in the elimination of light absorption by an optical media, electromagnetically induced transparency (EIT) \cite{HarrisPRLv641107,Fleischhauer2005RMP,Scully}. Scaling the media down to a single atom perfectly coupled to the incident waves leads to qualitatively new properties of EIT: the waves are scattered rather than absorbed. However, the strong interaction between the spatial electromagnetic modes and natural atoms (molecules, quantum dots) is very difficult to realize in practice \cite{Gerhardt,Wrigge,Tey,Vamivakas,Muller,MolAmp}. Recently, the strong ``atom''-field interaction has been achieved by confining the waves in the 1D transmission line efficiently coupled to an artificial three-level atom \cite{Astafiev2009} --- a superconducting quantum circuit. In a series of experiments \cite{Nakamura1999,Martinis2002,Chiorescu2003,Wallraff,Houck,Schuster,Fragner,Astafiev,Hofheinz} many fundamental quantum effects known from quantum optics, atomic physics and nuclear magnetic resonance have been reproduced using superconducting quantum circuits. However, most of those works focused on the two lowest levels. EIT related phenomena in superconducting circuits was theoretically studied in Ref.~\cite{Murali2004}.

In a few recent experiments multilevel structure of superconducting quantum circuits have been used to demonstrate Autler-Townes splitting and coherent population trapping \cite{Baur2009,Sillanpaa2009,Kelly2009}. Baur et al. observed Autler-Townes splitting in a three level quantum system coupled to a cavity using dispersive measurement. In Refs.~\cite{Sillanpaa2009,Kelly2009}, only level occupations were measured. In neither of these experiments direct transmission of probe field was measured.

Our artificial atom is a macroscopic size ($>$ 1 $\mu$m) superconducting loop interrupted by four Josephson junctions, which has been used as a two-level system or a flux qubit \cite{Mooij1999,Chiorescu2003} in earlier experiments. We exploit three lowest states $|i\rangle$ $(i= 1,2,3)$ with energies $\hbar\omega_{i}$, schematically represented in Fig.~\ref{fig1}a. The device parameters are designed such that all relevant transition frequencies of the three-level system $\omega_{ij} = \omega_i -\omega_j$ ($i>j$) fall within the frequency band of our experimental setup limited by 40 GHz. The atomic levels are controlled by the external magnetic flux thread through the loop $\Phi=\Phi_0/2+\delta\Phi$, where $\Phi_0$ is the flux quantum and $\delta\Phi$ is the deviation from $\Phi_0/2$. The loop is inductively coupled to a transmission line (open 1D space) via a mutual kinetic inductance $M$ \cite{Abdumalikov}. At the degeneracy point $\delta\Phi=0$ the atom has a ladder type energy level structure. Due to symmetry of eigenstate wavefunctions $|1\rangle\leftrightarrow|2\rangle$ and $|2\rangle\leftrightarrow|3\rangle$ transitions are allowed, while the transition $|1\rangle\leftrightarrow|3\rangle$ is forbidden. The coplanar transmission line with the characteristic impedance $Z\simeq50\,\Omega$ was made by patterning a gold film deposited on a silicon substrate. In the middle of the chip, the central conductor of the waveguide is narrowed and replaced by an aluminium strip. Our experiment is performed in a dilution refrigerator at a temperature of 40 mK.

In the rotating wave approximation, the three-level system under the drive of two fields with frequencies $\omega_{\rm p} = \omega_{21} + \delta\omega_{\rm p}$ and $\omega_{\rm c} = \omega_{32} + \delta\omega_{\rm c}$, where $\delta\omega_{\rm p}$ and $\delta\omega_{\rm c}$ presents small detunings from corresponding resonances $\omega_{21}$ and $\omega_{32}$, is described by the Hamiltonian
\begin{multline}\label{Eq1}
    H_a = -\hbar(\delta\omega_{\rm p} \sigma_{22} + (\delta\omega_{\rm p}+\delta\omega_{\rm c}) \sigma_{33}) - \\
    - \hbar\Bigg[\frac{\Omega_{\rm p}}{2}(\sigma_{21}+\sigma_{12}) + \frac{\Omega_{\rm c}}{2}(\sigma_{32}+\sigma_{32})\Bigg].
\end{multline}
Here $\sigma_{ij} = |i\rangle\langle j|$ is the atomic projection/transition operator, and $\hbar\Omega_{\rm p} = \phi_{21} I_{\rm p}$ and $\hbar\Omega_{\rm c} = \phi_{32} I_{\rm c}$ are the probe and control dipole interaction energies for the transitions $|2\rangle \leftrightarrow|1\rangle$ and $|3\rangle \leftrightarrow|2\rangle$, respectively. Under the influence of the two fields in the transmission line with the actual current values given by ${\rm{Re}}[I_{\rm p}(0,t)] = I_{\rm p}\cos{\omega_{\rm p} t}$ and  ${\rm{Re}}[I_{\rm c}(0,t)] = I_{\rm c} \cos{\omega_{\rm c} t}$. Here we assume that our pointlike atom is situated at $x=0$ and the waves from microwave sources $I_{\rm p}(x,t) = I_{\rm p}\exp{(ik_{\rm p} x-i\omega_{\rm p} t)}$ and $I_{\rm c}(x,t) = I_{\rm c}\exp{(ik_{\rm c} x-i\omega_{\rm c} t)}$ propagate in the transmission line.  The dipole matrix element can be presented in the form  $\phi_{ij} = \zeta_{ij}Mi_{\rm PC}$ with the dimensionality of a magnetic flux, where $\zeta_{ij}$ is the dimensionless matrix element ($0\leq \zeta_{ij}\leq 1$), $M$ is the line-atom mutual inductance, and $i_{\rm PC}$ is the amplitude of the persistent current in the loop.

The atomic dynamics is described by the Markovian master equation for the density matrix $\rho = \rho_{ij}|i\rangle\langle j|$,
\begin{equation}\label{Master}
\dot{\rho} = -(i/\hbar)[H,\rho]+L[\rho]
\end{equation}
with the Lindblad term
\begin{multline}\label{Eq3}
    L[\rho] = \Gamma_{32}\rho_{33}(-\sigma_{33}+\sigma_{22}) + \\ + \Gamma_{21}\rho_{22}(-\sigma_{22}+\sigma_{11})+ \sum_{i\neq j}\gamma_{ij}\rho_{ij}\sigma_{ij}.
\end{multline}
Here $\gamma_{ij} = \gamma_{ji}$ is the damping rate of the off-diagonal terms (dephasing) and $\Gamma_{ij}$ is the relaxation rate between the levels $|i\rangle$ and $|j\rangle$ ($i>j$). In the ladder-type three-level atom the $|3\rangle\rightarrow|1\rangle$ transition is omitted, since $\Gamma_{31}=0$. In our case the condition ($\hbar\omega_{ij}\gg k_BT$) is fulfilled and the absence of thermal excitations ($\Gamma_{12}$ = 0 and $\Gamma_{23}$ = 0) is guaranteed.

One can show that the atom interacting only with continuum modes of 1D open space generates a scattered wave at the probe frequency \cite{Astafiev2009}
\begin{equation}\label{Isc}
    I_{\rm sc}(x,t)=i\frac{\hbar\Gamma_{21}}{\phi_{21}}\langle\sigma_{12}\rangle e^{ik|x|-i\omega_{\rm p} t},
\end{equation}
where $\langle\sigma_{ij}\rangle= {\tt tr}[\sigma_{ij}\rho] = \rho_{ji}$ can be straightforwardly found in the stationary conditions ($\dot{\rho}=0$), when the master equation reduces to a set of linear algebraic equations. The transmission coefficient found as a ratio of the resulting current $I_{\rm p}(x,t) + I_{\rm sc}(x,t)$ (at $x>0$) to the incident one $I_{\rm p}(x,t)$ and for weak probe drive $\Omega_{\rm p} \ll \gamma_{21}$, the transmission coefficient is
\begin{equation}\label{trans}
    t = 1 - \frac{\Gamma_{21}}{2(\gamma_{21} - i\delta\omega_p)+\frac{\Omega_{\rm c}^2}{2(\gamma_{31}-i\delta\omega_{p}-i\delta\omega_{c})}}.
\end{equation}

The complex transmission coefficient $t$ is monitored using a vector network analyzer in the frequency range from 5\,GHz to 13\,GHz. Measuring $t$ versus probing frequency $\omega_{\rm p}$ and the magnetic flux $\delta\Phi$, the resonant transition frequency $\omega_{21}$ is revealed as a sharp dip in $|t|$ (dark line in Fig.~\ref{fig1}b). By fitting $\omega_{21}$ we find that the minimal frequency $\omega_{21}/2\pi$ reaches 10.165~GHz at $\delta\Phi=0$, and the persistent current in the loop is $i_{\rm PC} = 200$~nA. Next analyzing the spectroscopy line shape of $|t|$ at $\delta\Phi=0$, the relaxation and dephasing rates are found to be $\Gamma_{21} = 6.9\times10^7$ s$^{-1}$ ($\Gamma_{21}/2\pi=11$~MHz) and $\gamma_{21} = 4.5\times10^7$ s$^{-1}$ ($\gamma_{21}/2\pi=7.2$~MHz) \cite{Astafiev2009}. Non-radiative emission in such systems is expected to be negligible with corresponding relaxation rate less than $10^{-6}$ s$^{-1}$, measured in earlier experiments \cite{Yoshihara}. Therefore, we conclude that the relaxation is caused solely by the quantum noise of open 1D space defined by the rate $\Gamma_{21}=\hbar\omega_{21}(Mi_{\rm PC})^2/(\hbar^2Z)$ and derive the mutual inductance between the loop and the transmission line to be $M\approx12$~pH.

The transition frequencies $\omega_{31}$ and $\omega_{32}$ cannot be probed in the direct transmission, since they exceed the high frequency cutoff (13~GHz) of the cryogenic amplifier. To find them we use two-tone spectroscopy: The frequency of a weak probe tone, $\omega_{\rm p}$, is adjusted to $\omega_{21}$ (found from the data shown in Fig.~\ref{fig1}b), where the transmission $t_0$ is minimal ($|t_0| < 1$). Next, the transmission $t$ is continuously monitored, while the second (control) tone frequency $\omega_{\rm c}$ is swept. When $\omega_{\rm c}$ is in resonance with the corresponding transitions, total population  $|1\rangle$ and $|2\rangle$ is decreased and, therefore, $|t|$ is enhanced, revealing the spectral lines in plot of $|t/t_0|$. Fig.~\ref{fig1}c and Fig.~\ref{fig1}d show traces of the spectroscopy lines for $\omega_{32}$ and $\omega_{31}$, respectively. Due to symmetry of eigenstate wave functions, the selection rule in our system prohibits transitions between states $|1\rangle$ and $|3\rangle$ at $\delta\Phi=0$, which is seen in Fig.~\ref{fig1}c as the vanishing spectroscopy signal. At the same time, the matrix elements for $|1\rangle$ to $|2\rangle$ and $|2\rangle$ to $|3\rangle$ transitions reach their maximum. Thus at the degeneracy point, we have a ladder type three-level quantum system \cite{Scully}, schematically represented in Fig.~\ref{fig1}a, with two allowed and one forbidden transitions. The enhancement of transmission (suppression of reflection) in Fig.~\ref{fig1}c is already a signature of the single-atom EIT, which we study in detail at $\delta\Phi = 0$.

\begin{figure}
\includegraphics{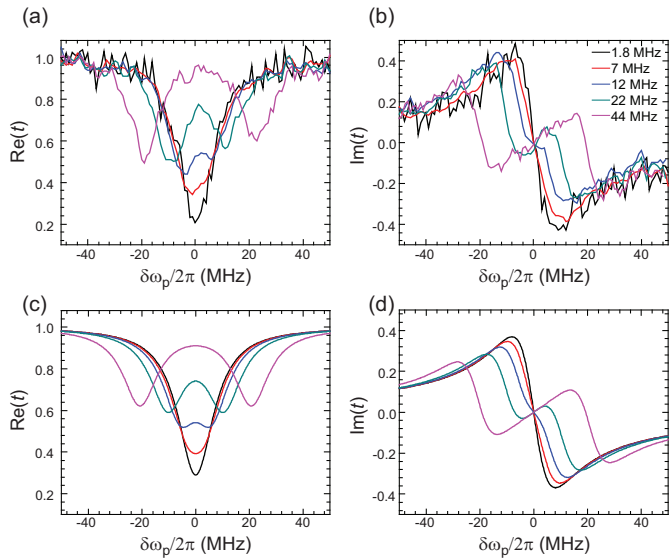}
\caption{\label{fig2} (Color online) Transmission coefficient near the probe wave resonance ($\delta\omega_p = 0$). (a) and (b) are real and imaginary parts of the probe signal transmission coefficient at the different control field amplitudes $\Omega_{\rm c}$, specified in (b). The curves show typical dispersion for EIT. (c) and (d) present calculations of real and imaginary parts for the probe signal transmission coefficient for the same set of $\Omega_{\rm c}$ as in (a) and (b).}
\end{figure}

Electromagnetic response of the single atom is naturally characterized by polarizability, $\alpha$ (rather than susceptibility used to characterize the optical response of macroscopic media). The polarizability is a ratio of the atomic dipole moment to the excitation field (in our case defined as the induced magnetic flux in the loop to the incident wave current amplitude). The atom scatters waves \cite{Astafiev2009} and the polarizability is related to the reflection and the transmission according to $\alpha \propto i r = i(1-t)$. The polarizability can be presented as $\alpha = \alpha' + i\alpha''$ with the real and imaginary parts  $\alpha'$ and $\alpha''$ related to dispersion and reflection, respectively. Fig.~\ref{fig2}a demonstrates transmission Re$(t) \propto 1-\alpha''$ for different amplitudes $\Omega_{\rm c}$ (associated with absorption in EIT for media), while Fig.~\ref{fig2}b shows dispersion curves of Im$(t) \propto \alpha'$. The probing amplitude in the measurements is fixed to $\Omega_{\rm p}/2\pi \simeq 2\,$MHz. In the absence of control field (black curves in Fig.~\ref{fig2}), the wave is strongly reflected, exhibiting the Lorentzian  dip in Re$(t)$, while Im$(t)$ follows the typical anomalous dispersion curve in vicinity of the resonant transition. With increasing control field amplitude $\Omega_{\rm c}$, the dip is split, and at the strongest drive ($\Omega_{\rm c}/2\pi$ = 44~MHz) the dip is completely suppressed, exhibiting full transparency at the exact resonance ($\delta\omega_{\rm p} = 0$) and a dispersion curve typical for EIT in Fig.~\ref{fig2}b. Figures~\ref{fig2}c and ~\ref{fig2}d show our calculations of  Re$(t)$ and Im$(t)$ respectively with $\gamma_{31}=4.3\times10^7$\,s$^{-1}$ ($\gamma_{31}/2\pi=6.9$\,MHz), which is comparable to $\gamma_{21}$. In atomic physics, the transmission window much narrower than the absorption dip appears already for weak control amplitude ($\Omega_{\rm c}\ll\gamma_{21}$) because of small dephasing between levels $|1\rangle$ and $|3\rangle$ $(\gamma_{31}\ll\gamma_{21})$ \cite{Fleischhauer2005RMP}.

\begin{figure}
\includegraphics[width=8cm]{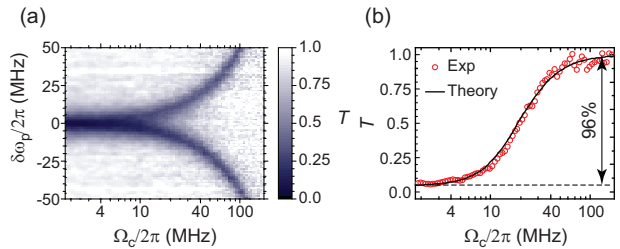}
\caption{\label{fig3} (Color online) Quantum switch operation. (a) Power transmission coefficient $T$ vs. control field amplitude $\Omega_c$ and probe frequency. The single resonant dip in the transmission is split into two under the large control field amplitudes. (b), $T$ for resonant probe signal vs. control field amplitude. The experimentally measured $T$ is presented by red circles, and the black line is calculated $|t|^2$ from Eq.~(\ref{trans}). The achieved contrast in the transmitted power is 96\%.}
\end{figure}

Figure~~\ref{fig3}a summarizes the data of Figs.~\ref{fig2}a-b, showing the power transmission coefficient $T = |t|^2$ as a function of control field amplitude $\Omega_{\rm c}$. The splitting at the strong drive is known as the Autler-Townes splitting \cite{AutlerTownes}. It arises due to Rabi splitting of levels $|2\rangle$ and $|3\rangle$. In the present experiments we could induce larger than 100 MHz splitting. Figures~\ref{fig2} and ~\ref{fig3}a demonstrate that the transmission strongly depends on the control field $\Omega_{\rm c}$, and, therefore, the latter can be used to control transmission and reflection for the probing wave. However, all the power can be reflected or transmitted only in extreme cases of $\Omega_c = 0$ or $\Omega_c \gg \gamma_{31}$, respectively. The power transmission $T$ at the exact probing wave resonance ($\delta\omega_{\rm p} = 0$) is presented in Fig.~\ref{fig3}b. The transmitted wave extinction exhibits contrast of 96\%, which demonstrates that the artificial atom can be used as a highly efficient directional switch (or mirror) for propagating waves. The power extinction is close to the ideal case of 100\%, which is possible in the absence of pure dephasing for the probe transition ($\gamma_{21} = \Gamma_{21}/2$) and if all the incident power interacts with the atom. In such a case the power transmission is presented by the simple formula
\begin{equation}\label{trans1}
T = \bigg(\frac{\Omega_{\rm c}^2}{2\Gamma_{21}\gamma_{31}+\Omega_{\rm c}^2}\bigg)^2.
\end{equation}
The black curve in Fig.~\ref{fig3}b shows calculation of $T$ from Eq. (\ref{trans}) as $|t|^2$ for our case of weak pure dephasing, which slightly deviates from Eq. (\ref{trans1}).

In conclusion, we have demonstrated operation of a quantum switch for propagating waves, which allows the propagating waves to be fully transmitted or backscattered. The experiment suggests interesting applications in photonics and optical quantum computation. It also demonstrates possibility of controlling individual atoms coupled to a 1D transmission line, which can be used, e.g., for photonic quantum information processing \cite{Kok}.



\end{document}